# Design of Transmission Line and Electromagnetic Field Sensors for DC Partial Discharge Analysis

**Mojtaba Rostaghi-Chalaki, Kamran Yousefpour, J. Patrick Donohoe, Mehmet Kurum** and **Chanyeop Park**[*]

Mississippi State University
Department of Electrical and Computer Engineering
Mississippi State, MS 39762, USA

**Joni Klüss**

Rise Research Institutes of Sweden
Measurement Science and Technology
Brinellgatan 4, Borås SE-501 15, Sweden

## ABSTRACT

**Accurate measurement circuit and high-frequency sensors with sufficient bandwidth are necessary for the analysis of individual partial discharge (PD) pulses. In this paper, a testbed is designed and constructed for the investigation of DC PD pulses. The testbed is equipped with a 50 Ω transmission line (TL) that terminate to an oscilloscope for measuring the charge displacement current generated by PD pulses. Besides the oscilloscope measurements, two types of electromagnetic field sensors (D-dot and B-dot) were developed to capture the EM fields of the PD pulses propagating through the TL. The main goal of this paper is to investigate the DC PD pulses through the EM fields and the corresponding discharge current pulses that are considered as calibrating signals for the developed D-dot and B-dot sensors. The results of DC cavity discharge measured by the constructed testbed and the EM field sensors demonstrate close agreement with the reference PD pulses measured via oscilloscope.**

## 1 INTRODUCTION

**DC** networks are growing rapidly in various industries (e.g. transmission and distribution grids, electric ships, aircraft, and power-electronics-driven systems) and becoming effective alternatives of conventional AC systems owing to their advantages including the capability of carrying more power with lower loss over long distances, the increasing availability and reducing cost of renewable energy sources, and the flexibility of converting energy among unsynchronized networks. However, the reliability of these power networks is threatened by the accelerated dielectric material aging and failure of subsystem components such as cables, transformers, switchgear, insulators, and bushings under DC electrical stress. Therefore, accurate and reliable assessments of these components are necessary to ensure the reliability of DC power networks. PD is one of the most chronic and unavoidable dielectric challenges that causes accelerated dielectric material aging and device failure. To name a few, the existence of metal particles in gas insulated substations (GIS) causes PD that lead to system failure by generating floating potentials, flashover, carbonization along spacers, and decomposition of gases [1, 2]. In the cases of cables, bushing, and transformers that are electrically insulated mainly by solid and liquid insulators, the inevitable formation of air bubbles (in liquid) or cavities (in solid) during the manufacturing processes leads to the generation of internal PDs [3, 4]. Hence, accurate PD measurement is integral for the condition assessment and health monitoring of medium- and high-voltage DC applications. However, the accurate measurement and interpretation of PDs under DC stress are more challenging than those of AC PDs that are accompanied by phase-resolved information. The phase-resolved partial discharge analysis (PRPDA) techniques are widely used for establishing patterns of various types of PDs (e.g. corona, cavity, and surface discharges) in AC systems. Advantages provided by PRPDA such as convenient PD interpretation and noise rejection are not available in analyzing DC networks. Therefore, utilizing individual PD pulse characteristics with respect to time, also known as time-resolved partial discharge (TRPD), is an option for PD interpretation under DC stresses. TRPD interpretations commonly include various methods of analysis according to the individual PD pulse characteristics versus their time of occurrences, the correlation among the succeeding and preceding pulses, and the repetition rate of PDs [1, 7]. Therefore, TRPD analysis for the identification of DC PD waveform requires accurate PD detecting instruments. These instruments vary by how PD pulses manifest themselves (e.g. discharge current pulses, electromagnetic (EM) waves, etc.) and are commonly divided into the subcategories of conventional and non-conventional sensors. The conventional PD sensors refer to the detecting instruments following the IEC 60270 criteria for bandwidth, upper, and lower frequencies while the non-conventional ones refer to sensors such as the EM field ones that measure propagated EM fields by PD signals in the range of 3 MHz to 3 GHz. Ultrasonic sensors capture acoustic waves generated by PD activities and optic sensors detect light emitted by PDs. [5, 6].

Generally, individual PD pulses under AC stresses are characterized by double exponential or Gaussian functions that have time characteristics on the range of hundreds of picoseconds to hundreds of nanoseconds [7, 8]. Consequently, the PD pulses of AC stresses have frequency components on the range of MHz to GHz [9, 10]. Although DC PD pulse analysis has been a rare research topic, in recent years, researchers have increasingly reported their studies on individual PD waveforms under DC stresses [11-16]. Their findings show that DC PD pulses consist of double exponential functions and that their temporal characteristics are similar to those of AC PD pulses. The findings indicate that individual DC PD pulses should be measured accurately via properly designed high-frequency measuring circuits and sensors with sufficiently large bandwidth.

Conventional sensors are not suitable for individual PD pulse analysis due to their low sensitivities and limited frequency responses. Conventional sensor sensitivity depends on the ratio between the capacitances of test objects and coupling capacitors. Their frequency responses range from tens of kHz to MHz that are suitable for identifying PD patterns, but not for individual waveform analysis [17, 18]. Alternatively, IEC standard 60270 recommended the application of high bandwidth instruments including oscilloscopes in combination with appropriate coupling sensors to measure the waveform or frequency spectrums of PD pulses [19]. Recently, the high frequency current transformer (HFCT) and non-conventional high-frequency coupling antennas that operate on the range of UHF have been increasingly used for measuring PDs in AC systems owing to their advantages such as superior sensitivity, broad frequency range and reduced external noise [8, 9, 17, 20]. The UHF antennas that analyze PD pulses by capturing their EM fields are good alternatives of conventional PD detection sensors. The EM fields radiated by PD pulses are originated by the transition of electrons within PD sources. Their frequency characteristics vary by the speed of electrons, recovery processes, and interruptions in discharge currents [10]. Thus, each type of PD shows distinguishable forms of EM fields due to differences in discharge mechanisms. EM field sensors such as D-dot and B-dot are suitable for fulfilling our goal of analyzing individual DC PD pulses within an ideal transmission line. The D-dots and B-dots are capacitively- and inductively-coupled EM field sensors, respectively, which has low price, simple design, compact size, and very high frequency response ranges [21]. Also, convenience in the calibration process of the D-dot and B-dot sensors and their sensitivities comparable with the existing EM sensors make them more suitable for the measurement of transient voltages and currents. Using the D-dot and B-dot sensors to measure pulses with pico-seconds rising times in pulsed power applications have been reported in [21, 22]. Also, in [23], the D-dot sensors have been employed to detect multiple PD types such as corona, surface, and cavity discharges in switchgear and power cables. The applied D-dot sensors in [23] and existing similar researches include regular BNC and SMA connectors equipped with a ground plane which are improperly designed for PD measurements – especially not suitable for individual PD pulse measurements, the main goal of this work. Therefore, in this paper, the D-dot and B-dot design parameters that enable accurate individual PD pulse measurements are discussed. Generally, dimensions and locations of D-dot and B-dot sensors are critical for their sensitivity and bandwidth.

In addition to the sensors, impedance matching among circuit connections and data acquisition (DAQ) systems as well as providing a waveguide for the emitted EM fields of PD is essential to guarantee accurate individual PD pulse measurements. Application of UHF sensors in GIS showed high-order propagation modes of transverse electromagnetic (TEM) waves related to coaxial geometries [17, 20]. Therefore, designing a coaxial TL with suitable bandwidth and connecting it to coaxial cables provides a consistent 50 $\Omega$ impedance path for PD pulses and their EM waves traveling toward DAQ systems and D-dot and B-dot sensors. Poorly designed TL causes impedance mismatch among measurement circuit elements that results in wave reflections and signal losses [18]. As reported in [18], poorly designed TL causes significant wave reflections that lead to the deformation of original PD signals.

This paper aims to present the design requirements of testbeds used for the analysis of individual PD pulses under DC stresses based on non-convention PD sensors. First, all IEC 60270 requirements for PD measurement circuit are taken into account and designed. Subsequently, the detailed dimensions of the testbed are determined through finite element analysis (FEA) using Comsol Multiphysics. In addition, the specifications of designing suitable UHF sensors (D-dot and B-dot) for capturing the EM waves of PD pulses are explained. Furthermore, a 50 $\Omega$ coaxial TL that provides a reflection-free path for PD pulse-induced current and EM waves, is designed and constructed.

## 2  DESIGN AND ANALYSIS

This section focuses on the design of testbed equipped with a coaxial TL and non-conventional EM field sensors that accurately captures individual PD pulses. The frequency responses of the testbed, TL, and EM field sensors are numerically analyzed in Comsol Multiphysics. The numerical analysis tool is utilized to determine specific dimensions that enable matched impedance of the testbed and to accurately position the EM sensors in the TL.

### 2.1 TESTBED OVERVIEW

Figure 1a shows the AC PD measurement circuit suggested by IEC standard 60270 [19]. The PD measurement circuit includes an HV source (U), filter (Z), coupling capacitor ($C_k$), test object ($C_a$), coupling device (CD with input impedance $Z_{mi}$), connecting cable (CC), and measuring instrument (MI). Due to the lack of standards for DC PD measurement, recommended configurations of a high bandwidth testbed that measures individual PD pulses reported by Klueter *et.al* in [11] is used in this work. The testbed is shown in Figure 1c. As shown in the figure, the testbed consists of a cap that connects to a high voltage supply, a 9.86 pF air filled coupling capacitor between the HV cap and the shield electrode that isolates the PD sources from external noise and disturbances (ground electrode), and a matched 50 Ω conical TL (coaxial configuration) that terminates the PD source (Figure 1b) to an N-Type which grounds the entire testbed through the oscilloscope. In addition, PD sources are mounted between the coupling capacitor and the center electrode of the conical TL. The coaxial symmetry of the testbed design simplifies simulation and machining. For the accurate design of the testbed, all parts are simulated in Comsol Multiphysics to finalize the dimensions. Through the simulation, required clearances are determined and electric field distributions are analyzed to ensure corona-free condition. It should be noted that the maximum size of the testbed was limited to 90 mm, the maximum machinable size of lathe machine.

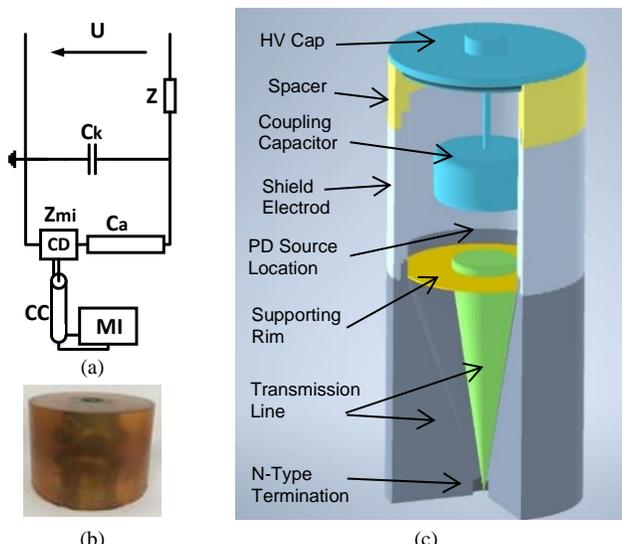

**Figure 1.** PD measurement circuit. a) Recommended circuit by IEC standard 60270 [19], b) Prepared cavity discharge source, which is placed between the coupling capacitor and TL, c) Cutaway view of the designed testbed for the investigation of individual PD pulses.

To verify the performance of the testbed, a case, in which a needle is connected to the coupling capacitor facing towards the center electrode of the TL is studied. Applying high voltage DC to the HV cap connected to the needle increases the electric field at the tip of the needle. As the voltage exceeds the required voltage of breakdown, after elapsing stochastic time lags (describe in [7]) corona PD occurs on the tip of the needle. Subsequently, the displacement of electrons in the gap between the needle tip and the center electrode of the TL flows a current through the conical TL and received by the oscilloscope. The discharge current generated by PD is measured by the 50 Ω input channel of the oscilloscope that is essentially a shunt resistor and reference for evaluating the performance of D-dot and B-dot sensors. The measurements of the D-dot and B-dot are the differentiated signals of the propagated E-field and H-field generated by PD pulses, respectively. It should note that the needle is only an example used to verify how the testbed works. The main PD type investigated in this work is a cavity discharge (see section 3.3, and Figure 1b).

Figure 2 represents the 2D axisymmetric simulation results of electric field and voltage distributions in the testbed. In this case, a needle is connected at the bottom of coupling capacitor. The applied voltage of 20 kV is evenly distributed as shown in Figure 2b among the high voltage cap, coupling capacitor, and grounded shield electrodes. As shown by Figure 2a, blue colors demonstrate the areas with the E-fields below 1 kV/mm. In addition, E-fields are strong around sharp edges, corners, and the interfaces among spacer and metal parts. Hence, the E-fields are reduced by applying larger radii on the edges to achieve values under 3 kV/mm, the breakdown field of atmospheric pressure air, to avoid corona discharge. The highest E-field values of each part in testbed are shown in Figure 2a at 20 kV.

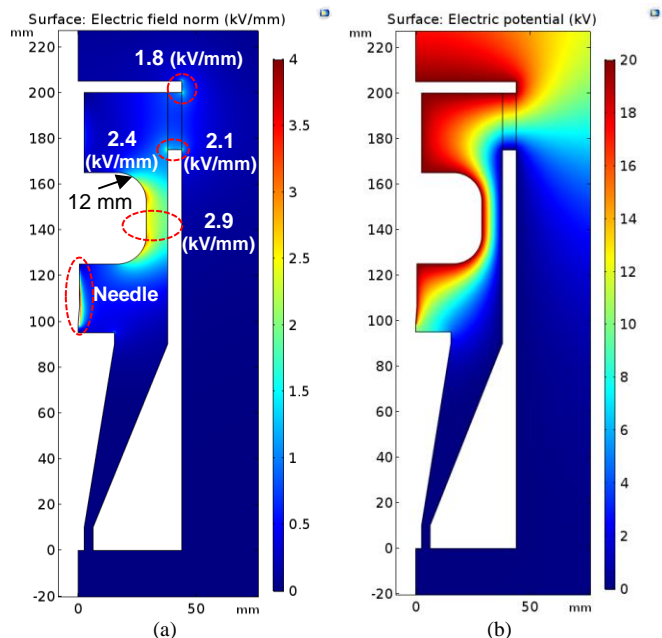

**Figure 2.** 2D axisymmetric simulation of DC PD testbed. a) Electric field distribution (kV/mm), b) Voltage distribution (kV).

## 2.2 TRANSMISSION LINE (TL)

TLs are widely used in the measurement circuits of transient and high-frequency phenomena. Various types of TL work as wave guiding structures that provide paths for signals to travel with minimal reflection and impedance mismatch. In the case of individual PD pulse measurement, it is important to connect the PD source to a measurement device without external disturbances. To ensure the noise-free condition, a TL that has impedance matched with all components of the measurement apparatus (e.g. connectors, coaxial cables) and the DAQ system is designed. As illustrated in Figure 1, the best configuration for terminating the PD source to the measuring circuit is using a conical TL. The upper surface of the conical TL provides enough planner surface for the PD source installation (e.g. the needle-plane configuration which simulates PD at sharp tips). The tapered diameter of the conical TL enables proper PD signal termination to RF connectors such as N-type and UHF connectors at the base of the testbed.

Coaxial cables are the most common and simple form of TL that are available in a wide range of frequencies and suitable for various measuring circuits. These cables consist of layers of electrodes that improve shielding as shown in Figure 3a. Also, the coaxial configuration provides a uniform area for the propagation of electric and magnetic fields within the TL which are isolated from external disturbances. The most important parameter that should always be considered in the design of a coaxial TL is its characteristic impedance. Many measurement circuits are designed for 50 Ω impedance for specific ranges of frequencies and its critical to design such a TL with a 50 Ω impedance.

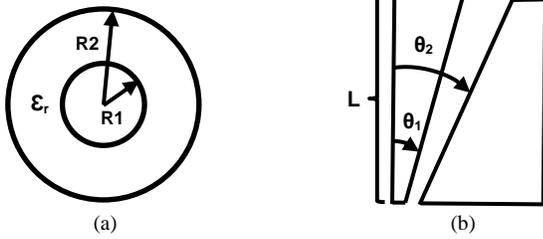

**Figure 3.** Schematics of coaxial TLs. a) Cylindrical TL, b) Conical TL.

Generally, the characteristic impedance ($Z_0$) of a cylindrical coaxial TL (Figure 3a) is expressed by Equation (1) [24],

$$Z_0 = \left(\frac{1}{2\pi}\right)\left(\sqrt{\mu/\varepsilon}\right) ln\left(\frac{R_2}{R_1}\right) \approx \left(\frac{60}{\sqrt{\varepsilon_r}}\right) ln\left(\frac{R_2}{R_1}\right) \quad (1)$$

where, $\mu$ and $\varepsilon$ are the permeability and permittivity, respectively, of the nonmagnetic insulating material between the conductors ($\varepsilon_r$ is the relative permittivity). Also, $R_2$ is the radius of the outer electrode and $R_1$ is the radius of the center electrode. Similarly, the characteristic impedance of a conical TL illustrated in Figure 3b is shown in Equation (2) [24],

$$Z_0 = \left(\frac{60}{\sqrt{\varepsilon_r}}\right) ln\left(tan\left(\theta_2/2\right)/tan\left(\theta_1/2\right)\right) \quad (2)$$

where $\theta_2$ and $\theta_1$ are the outer and inner electrode angles, respectively. Equation (2) shows that $Z_0$ is determined by the ratio between center and outer electrode angles ($\theta_1$ and $\theta_2$) of the conical TL.

A numerical analysis is performed to design the conical TL while accounting for the manufacturing limitations including the size of the lathe machine chuck and the length of boring lathe tool. The main purpose of the numerical analysis is examining the effect of the TL dimensions on its matching characteristics. Figure 4 shows the FEA results of Comsol Radio Frequency (RF) module. To decrease computational effort, parametric simulations are performed in 2D axisymmetric models. Moreover, to excite the TL, a coaxial 50 Ω lumped-port node was used to determine the input and output terminals with impedances matched to 50 Ω. To evaluate the frequency response of the TL, S-parameters such as S11 and S21 were used to measure reflections and losses along the TL. S11 is forward reflection coefficient (input match) at port 1 and S21 is forward transmission coefficient (gain or loss) from port 1 to port 2 when port 2 terminated in a perfect match.

Figure 4a shows the top view of electric field (red arrows) and magnetic field (green arrows) directions and surface current density at the sample frequency of 1 GHz. According to this figure, the simulated TL design provides a proper path for the EM fields radiated by the PD displacement currents that are captured by the D-dot and B-dot sensors. Also, Figure 4b and 5c illustrate the result of parametric sweep simulations. The simulation includes the sweeping of $\theta_1$ and length of TL (L), and the value of $\theta_2$ is calculated through Equation (2) when the $Z_0$ is equal to 50 Ω. According to the results, increasing $\theta_1$ causes more reflections and greater losses while increasing TL's length reduces both reflections and losses. Based on machining limitations, the selected conical TL specifications are θ1 = 15º, θ2 = 33.7º and L = 80 mm that provides low reflection and high cut-off frequency (higher than 6 GHz).

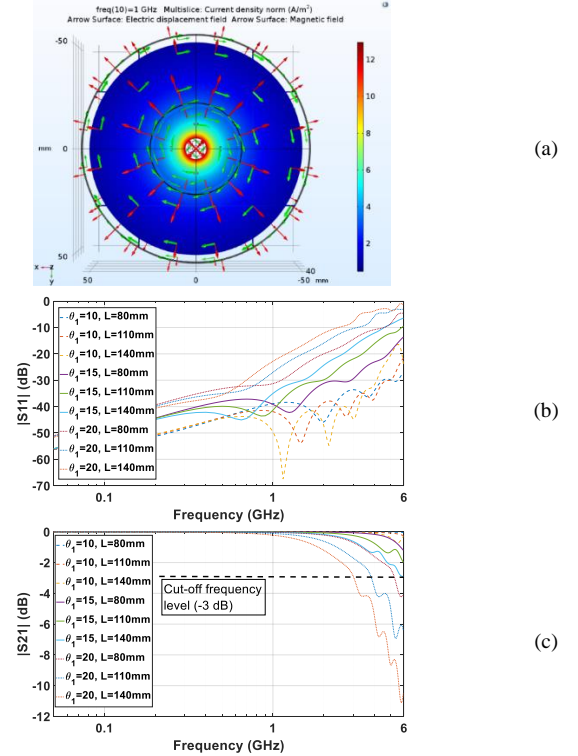

**Figure 4.** FEA results of conical transmission lines. a) EM field: red and green arrows are E-fields and H-field, respectively, b) S11: reflection coefficient, c) S21: forward transmission.

## 2.3 ELECTROMAGNETIC FIELD SENSORS

The non-conventional EM field sensors are a set of inductive (e.g. loop antenna) or capacitive (e.g. monopole antenna) coupling sensors that capture EM waves and react to very fast transient phenomenon in the HF/VHF/UHF ranges. To clarify, non-conventional PD sensors are those that are not installed within the current path of PDs unlike the conventional PD sensors that directly measure PD current. In this study, two differentiating type non-conventional sensors, D-dots and B-dots, are developed to respectively measure the electric and magnetic fields radiated from the individual PD pulses. Figure 5 shows the location of sensors within the conical TL, their equivalent circuits, and their schematics. The dimension and location of sensors were determined by both FEA simulation and experiment. During the FEA simulation, the dimensions of the sensors were determined by accounting for the criteria of Equations (3) and (5) that are related to the differentiating mode of sensors, which will explain in detail later. Also, the mounting location of the sensors was determined experimentally by selecting various locations along the shielding electrode of TL to find the sites with sufficient sensitivity.

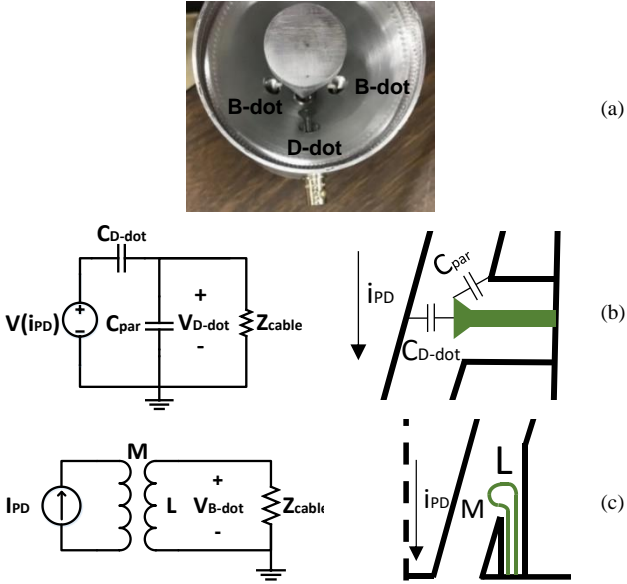

**Figure 5.** Principles of non-conventional sensors. a) D-dot and B-dot sensors mounted in the TL, b) Equivalent circuit and schematic of D-dot [21, 22], c) Equivalent circuit and schematic of B-dot [21, 22].

As shown in Figure 5a, there are two B-dots (loop antenna) and a D-dot (flat head monopole antenna) sensors installed at the bottom and middle of TL, respectively. To utilize high-intensity magnetic fields occurring at the bottom end of the conical TL, where it is terminated to an N-type connector, the B-dot sensors were placed near the bottom. Here two B-dot sensors are utilized to suppress common mode noise [22]. Figure 5b shows the equivalent circuit and schematic of a D-dot sensor. As shown, the D-dot sensor is capacitively coupled via $C_{D\text{-}dot}$ and $C_{par}$ with the center electrode of the TL and the shielding of TL, respectively. The parasitic capacitances ($C_{par}$) are in parallel with the characteristic impedances of connecting cable ($Z_{cable}$). Therefore, induced voltage by the discharge current of individual PD ($V(i_{PD})$) is calculated by Equation (3) [22].

$$V(i_{PD}) = \left(\frac{C_{Ddot} + C_{par}}{C_{D-dot}}\right)V_{Ddot} + \frac{1}{Z_{cable}C_{Ddot}}\int_0^t V_{Ddot}dt \quad (3)$$

Consequently, by converting Equation (3) into frequency domain, the transfer function of D-dot sensor ($H_{D\text{-}dot}(j\omega)$) is obtained as in Equation (4) [22].

$$H_{Ddot}(j\omega) = \frac{V_{Ddot}(j\omega)}{V_{iPD}(j\omega)} = \frac{j\omega C_{Ddot}Z_{cable}}{j\omega(C_{Ddot} + C_{par}) + 1} \quad (4)$$

Equation (4) shows that the D-dot sensor operates as a differentiating sensor for frequencies that satisfy $j\omega(C_{Ddot} + C_{par}) \ll 1$. On the other hand, the sensor operates in a proportional mode (*i.e.* self-integrating mode) for frequencies that satisfy $j\omega(C_{Ddot} + C_{par}) \gg 1$. Generally, the size of sensor tip (flat head area) and the distance of sensor from the center electrode of TL (location) determines the value of $C_{D\text{-}dot}$.

Accordingly, the $C_{D\text{-}dot}$ of Equation (4) determines the sensitivity and upper-band frequency of differentiating mode.

The model of B-dot in Figure 5c consists of a mutual inductance *M* between the sensors and the TL's center electrode and a self-inductance *L* of the sensor. Therefore, individual PD pulse current can be calculated through Equation (5) [22].

$$I_{PD} = \frac{1}{M}\left(\frac{L}{Z_{cable}}V_{Bdot} + \int_0^t V_{Bdot}dt\right) \quad (5)$$

Subsequently, the transfer function of the B-dot sensor is modeled as follows [22].

$$H_{Bdot}(j\omega) = \frac{V_{Bdot}(j\omega)}{I_{PD}(j\omega)} = \frac{j\omega M}{j\omega\frac{L}{Z_{cable}} + 1} \quad (6)$$

Equation (6) shows that the B-dot sensor performs in a differentiating mode at frequencies that satisfy $j\omega\frac{L}{Z_{cable}} \ll 1$. On the other hand, it works in a proportional mode for frequencies that satisfy $j\omega\frac{L}{Z_{cable}} \gg 1$. The value of M in Equation (6) is a function of distance between the B-dot loop and the center electrode of the TL. Also, the size of the B-dot loop and wire thickness determine the value of L. By adjusting the M and L values, the sensitivity and upper-band frequency of the differentiating mode can be controlled. However, it should be noted, the sensors should be operated in either of the two modes (differentiating or proportional) within a given bandwidth to enable consistency in the distinguishing process of signals.

## 3 RESULTS AND DISCUSSION

In this section, the results of frequency response and bandwidth of the conical TL and the D-dot and B-dot sensors are presented. Also, the individual PD pulses captured using testbed are compared with the reference pulses.

### 3.1 Transmission Line Frequency Response

To measure the s-parameters of the TL, two conical TLs are connected back to back such that both ends are connected to a network analyzer (NA). Figure 6 shows the TLs connected to a NA and the frequency responses (S11 and S21) of TL. The NA used is KEYSIGHT FieldFox Analyzer N9917A.

Figure 6a shows the experimental setup of the frequency response measurement. Ports 1 and 2 of the NA are connected to both ends of the TL. Figure 6b shows the magnitude of reflection (S11) and losses (S21) versus frequency. Generally, S11 values less than -10 dB are considered satisfactory. As shown in this figure, the designed conical TL's reflection is less than -10 dB up to 2.2 GHz. Moreover, the cut-off frequency of S21 of TL is higher than 3 GHz. The obtained results are not exactly in accordance with the simulation results in Figure 4b and 4c due to the differences between the ideal assumptions of the simulation (purely 50 Ω coaxial lumped port) and the actual experimental setup. However, measurements shown in Figure 6b suggest that the designed TL has a sufficiently wide bandwidth and low reflections and attenuations, thus suitable for individual pulse measurements of PDs with dominant frequencies lower than 2.2 GHz.

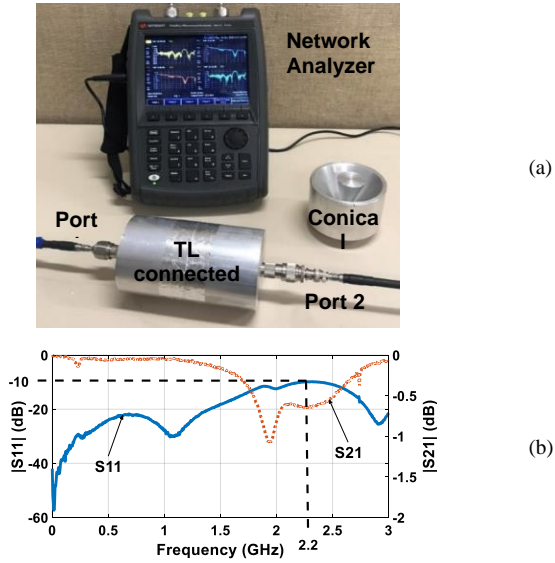

**Figure 6.** S-parameter measurement setup of the designed TL.
a) Test circuit, b) S-parameters (S11 and S21).
*Note:* The graphs shown in the screen of the NA are mere examples.

### 3.2 D-dot and B-dot Frequency Response

The frequency responses of the designed D-dot and B-dot sensors are shown in Figure 7. The testbed for measuring the sensor frequency responses is identical to that shown in Figure 6a except that one end of the TL is connected to a 50 Ω load. Here, Port 1 of the NA is connected to the TL to excite power into the TL while Port 2 measures the frequency responses of the D-dot and B-dot sensors individually. Note that the differentiating mode refers to regions, in which the sensitivity of the D-dot and B-dot sensors change linearly in log scale while the proportional mode indicates constant sensitivity (e.g. shunt resistors, and HFCTs). According to Figure 7, the D-dot and B-dot sensors operate in a differentiating mode (described in 2.3) up to 1.2 and 1.4 GHz, respectively, and convert to proportional mode thereafter. In other words, the sensitivity of the sensors is a function of the frequency components of the observed signal. It improves with the increase of the frequency within the differentiating mode regions of Figure 7. Therefore, this confirms that the designed D-dot and B-dot sensors can measure individual PD pulses with rise times on the range of hundreds of picoseconds (*i.e.*, <1.2 GHz). Indeed, the frequency component and the amplitude of PD signals determine the sensitivity of the sensors according to Figure 7. In general, if the PD pulse has frequency components in the differentiating mode, the sensors outputs are the derivative of that. In general, measurement accuracies of the D-dot and B-dot sensors decrease within transition between two modes, due to the complexity of extracting original signals from sensor outputs [21, 22].

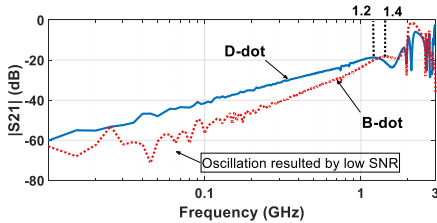

**Figure 7.** Frequency responses of the D-dot and B-dot sensors.

Therefore, it is recommended to operate sensors in either mode, but preferably in the differentiating mode due to simpler design. Moreover, it should be noted that the oscillations of the B-dot response below 0.1 GHz in Figure 7 are caused by the low signal to noise ratio (SNR) [21].

### 3.3 Partial Discharge measurement

To examine the performance of the designed testbed, which includes the 2.2 GHz bandwidth TL and the non-conventional sensors, PD pulses occurring in a dielectric sample are measured. The results of the cavity PD measurements are shown in Figure 8. Figure 8a shows the experimental setup used for measuring the PD pulses. The experimental setup includes a 200 kV DC Haefely Multi Test Set, a Haefely capacitive voltage divider, the designed PD testbed. Also, all the PD signals including reference, D-dot, and B-dot outputs were measured by the Teledyne LeCroy WaveMaster 806Zi-B (6 GHz - 4×40 GS/s) oscilloscope.

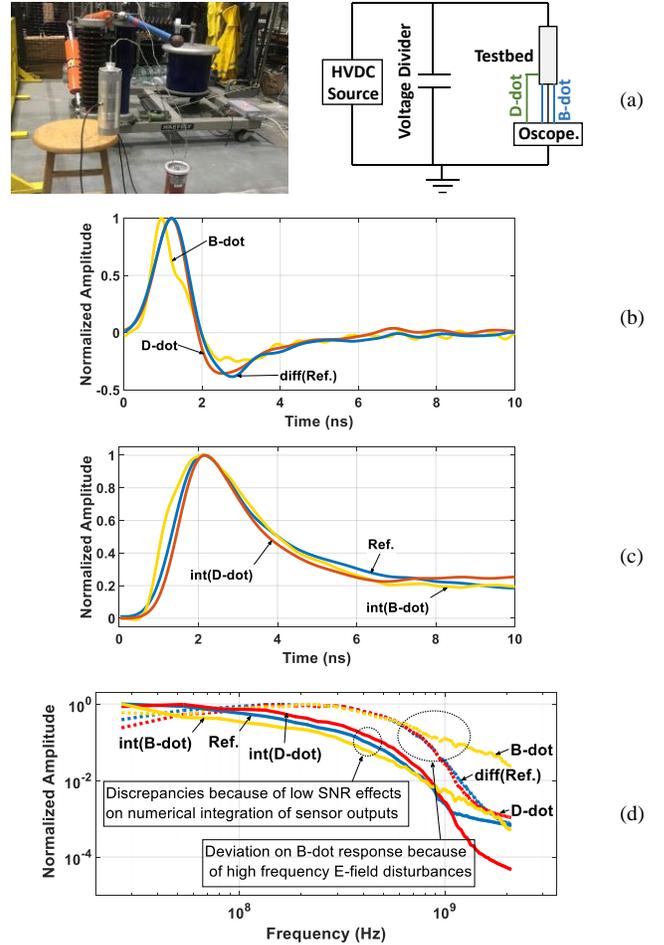

**Figure 8.** Experimental setup and the measured results of a cavity PD pulse. a) Test setup, b) EM sensor (B-dot and D-dot) responses, c) EM sensor response compared to the ref. pulse, d) FFT analysis of sensors.

The input 50 Ω resistance of the oscilloscope is considered as a reference signal for the calibration of both D-dot and B-dot sensors. One thing to note is that the benefit of using the D-dot and B-dot sensors for the investigation of the individual PD pulses is the protection of the oscilloscopes. Indeed, direct measurement of PD pulse through the 50 Ω channel of an oscilloscope causes the full discharge of PD energy to flow through the oscilloscope which could damage the scope channel in the case of strong PD pulses. Consequently, using EM field sensors makes a wireless connection between the PD source and oscilloscope for measuring the individual PD pulses. In the other words, after calibrating the sensors with the reference signal (50 Ω channel of oscilloscope), the testbed can be terminated with an external 50 Ω load instead of oscilloscope and the propagated EM fields by PD pulses can be captured by the D-dot and B-dot sensors. However, Figure 8b shows the normalized output of the B-dot and D-dot sensors as well as the differentiation of reference pulse which is measured by the 50 Ω input channel of an oscilloscope.

The figure clearly shows that the output signals of both EM sensors are the differentiation of PD pulses that are double exponential functions. Accordingly, the integration of the EM sensor outputs results in the reference PD waveform in time domain. Figure 8c compares the normalized integration of D-dot and B-dot output signals with the reference pulse. As shown, the outputs of the sensors fit well with the reference pulse, which contains rising time, decay time, 50% pulse width, 20% pulse width. Beyond 10 ns, the integrated signals deviate from the reference PD pulse due to the low SNR of the non-conventional sensors caused by the low $di/dt$ of the PD pulse. Table 1 presents statistical data comparing the integrated signals of D-dot and B-dot sensor measurements for 100 reference PD pulses. The table compares four characteristics of a double exponential waveform: rising time (Tr), decay time (Td), 50% and 20% pulse widths (PW50 and PW20, respectively). As shown in table, averages and standard deviations of both D-dot and B-dot sensor measurements are in agreement with those of the reference PD pulses in time domain. The low standard deviation among the recorded data for each sensor and the low error percentage of the D-dot and B-dot outputs compared to that of the reference confirms that design is suitable for the individual PD pulse investigations. Furthermore, the Fast Fourier Transform (FFT) of 100 PD pulses is performed to compare the frequency spectra of the designed D-dot and B-dot sensors to those of the reference as shown in Figure 8d. The FFT results show correlations among the three sensors: 50 Ω input resistance of oscilloscope (conventional), D-dot and B-dot (non-conventional). According to the results, the FFT response of the B-dot sensor begins to deviate from the reference signal at frequencies above 900 MHz. The substantial discrepancy shown in the high frequency range is mainly due to the high-frequency E-fields that cause disturbances on the response of the B-dot sensor. To suppress these disturbances, electrostatic shielding or analog filtering could be applied [22]. However, since the dominant frequency component of PD pulses are below 900 MHz, no filter was used for the B-dot measurements in this work. Comparing sensor performances in frequency domain shows that the differentiating mode (dotted lines) provides more accurate DC PD pulse analysis as shown in Figure 8d. The relatively poor performance of the B-dot sensor in the integrating mode (solid lines) is due to the integration of high frequency noise of the sensor caused by E-field disturbances. Figure 8d also shows that, in general, the performance of D-dot

(E-field) sensor is better than that of the B-dot sensor. Hence, if timing parameters are not of interest, the differentiating mode should be used for DC PD analysis since it does not introduce numerical integration errors.

In general, the time domain results suggest that the developed non-conventional sensors are capturing the transients of individual PD pulses well. The strength of EM fields radiated by individual PD pulse greatly depends on the rate of change of induced current [25]. For example, in the rising and falling regions of PD pulses (Figure 8c), where the $di/dt$ is relatively high, the D-dot and B-dot sensors performs well while their performance is comparatively poor in regions where pulses slowly decay. Therefore, the two non-conventional sensors (D-dot and B-dot) are reliable for the measurements of pulse characteristics including rising time, decay time, 50% pulse width, and 20% pulse width, all of which occur within the 10 ns range. The statistical data of these characteristic values enable the identification of various types of DC PD. However, the designed D-dot and B-dot sensors in this work are suitable for individual PD pulse waveshape measurement while they can be designed in the future works to be more applicable for the practical applications such as GIS substation PD measurements that deals with detecting mobile particle movements.

**Table 1.** D-dot and B-dot pulse characteristics vs reference pulse.

|      | Reference |      | D-dot         |              | B-dot         |              |
|------|-----------|------|---------------|--------------|---------------|--------------|
|      | Mean      | σ    | Mean (% error)| σ (% error)  | Mean (% error)| σ (% error)  |
| Tr   | 1.08      | 0.20 | 0.99 (% 8)    | 0.18 (% 10)  | 0.94 (% 13)   | 0.21 (% 5)   |
| Td   | 3.83      | 0.74 | 3.61 (% 6)    | 0.64 (% 13)  | 4.52 (% 18)   | 1.02 (% 38)  |
| PW50 | 2.72      | 0.59 | 2.19 (% 19)   | 0.45 (% 24)  | 3.45 (% 27)   | 1.12 (% 90)  |
| PW20 | 6.53      | 1.21 | 5.45 (% 16)   | 1.87 (% 54)  | 5.94 (% 9)    | 2.35 (% 94)  |

Note: All values are in nanoseconds (ns)

## 4 CONCLUSIONS

A testbed incorporated with an ideal TL and EM field sensors (D-dot and B-dot) was designed in this work. The designed testbed ensures a PD-free environment up to 20 kV and is equipped with a TL that has a bandwidth of 2.2 GHz. Designed TL provided low-loss and low-reflection path for PD pulses. Both D-dot and B-dot sensors were designed to capture EM fields radiated by individual PD pulses. The frequency responses of the sensors suggested that both D-dot and B-dot operate in a differentiating mode up to 1.2 and 1.4 GHz, respectively. The comparison of the D-dot and B-dot sensor outputs and the reference PD pulse showed close agreements in both time and frequency domains. The findings suggest that the conventional sensor as well as the electromagnetic field non-conventional sensors (D-dot and B-dot) are suitable for studying pulse characteristics including rising time, decay time, 50% pulse width, and 20% pulse width measured through TL where wave propagation characteristics are known. This can be used for investigating PD characteristics at the source and allow for further development of systems intended for onsite applications. Furthermore, the statistical data of pulse characteristics will be used for developing the identification methods of various types of DC PDs in future studies.

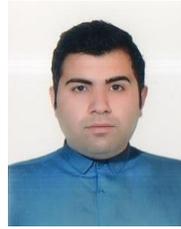

**Mojtaba Rostaghi** was born in Gorgan, Iran, in 1988. He received the B.S. from the SBU, and the M.S. from University of Tehran, Iran, in 2011 and 2013, respectively. He has joined the HVLab research group at MSU since 2017 and currently is a PhD candidate in Electrical Engineering. His research interests include the HV engineering and test techniques, PD measurement, HV apparatus risk assessment, electromagnetic field measurement, and Transient phenomenon.

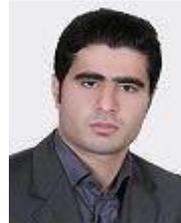

**Kamran Yousefpour** was born in Babol, Iran, in 1988. He received the B.S from Mazandaran University, Iran, in 2011, and M.S. degree from the IAU, Iran, in 2014. He is currently pursuing his PhD degree in Power Systems and High Voltage Engineering in HVLab at MSU from 2017. His research interests include transient and lightning impulse tests, investigation of carbon fiber composites subjected to lightning, FEA simulation of lightning, HV testing, and PD measurement.

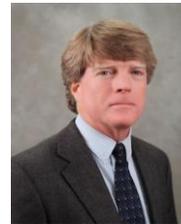

**J. Patrick Donohoe** received the B.S. and M.S. degrees in Electrical Engineering from MSU in 1980 and 1982, respectively. He received the Ph.D. degree in Electrical Engineering from the University of Mississippi in 1987. Dr. Donohoe began his career at MSU as an Assistant Professor of Electrical Engineering in 1986 and currently holds the rank of Professor and Paul B. Jacob Chair of Electrical and Computer Engineering at MSU. His primary research interests include computational electromagnetics, antennas, radar, electromagnetic compatibility, electromagnetic measurements, electromagnetic properties of composite materials, and lightning protection.

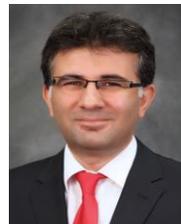

**Mehmet Kurum** (M'08–SM'14) received the B.S. degree in electrical and electronics engineering from Bogazici University, Turkey, in 2003, and the M.S. and Ph.D. degrees in Electrical Engineering from George Washington University, USA, in 2005 and 2009, respectively. He held a Postdoctoral position with the Hydrological Sciences Laboratory, NASA Goddard Space Flight Center, and Greenbelt, USA. In 2016, he was an Assistant Professor with the Dept. of Electrical and Computer Engineering, MSU. His research interests include High-resolution Earth imaging, Signals of Opportunity, Smartphone/Radar Sensing, Sensor fusion/ Machine Learning for Inverse problems.

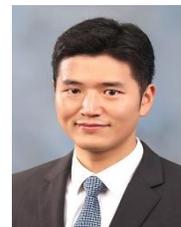

**Chanyeop Park** (M'19) is an Assistant Professor in the Dept. of Electrical and Computer Eng. at MSU. He is directing research activities in the Paul B. Jacob HVLab at MSU. Prior to joining MSU, he was a Postdoctoral Fellow at the Georgia Institute of Technology where he received his Ph.D. in Electrical and Computer Engineering in 2018. He received his M.S. and B.S. in Electrical Engineering from Hanyang University, Korea Rep. in 2013 and 2011, respectively.

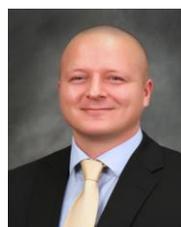

**Joni V. Kluss** received the M.S. degree from the Helsinki University of Technology, Finland, in 2008, and the PhD degree in Power Systems and HV Eng. from Aalto University, Finland, in 2011. He is currently a senior researcher in Measurement Science and Technology (HV Group) at RISE, Sweden. Prior to RISE, he was the director of the HVLab and assistant professor in the Dept. of Electrical Engineering at MSU, USA. His research interests span the fields of HV engineering ranging from gaseous insulation and dielectric breakdown to HV measurement techniques and EMP research.